\documentclass{trbunofficial}   
\usepackage{graphicx}
\usepackage{newtxtext,newtxmath}
\usepackage{bm}

\usepackage{float}

\usepackage[hidelinks]{hyperref}

\newif\ifarxiv
\arxivtrue      

\AuthorHeaders{Ahmad, Rahman, Sevim, Bodoh, Khan, Zhao, Huynh, and Ozguven}

\title{Historical Prediction Attention Mechanism based Trajectory Forecasting for Proactive Work Zone Safety in a Digital Twin Environment}
\author{
\textbf{Minhaj Uddin Ahmad, Ph.D. Student}\\
  Department of Civil, Construction \& Environmental Engineering, The University of Alabama\\
  Smart Communities and Innovation Building (SCIB), 28 Kirkbride Lane, 
  Tuscaloosa, AL 35487-0288\\
  Email: mahmad12@crimson.ua.edu \\
  \hfill\break 
    \textbf{Mizanur Rahman, Ph.D.}\\
  Department of Civil, Construction \& Environmental Engineering, The University of Alabama\\
  Smart Communities and Innovation Building (SCIB), 28 Kirkbride Lane, 
  Tuscaloosa, AL 35487-0288\\
  Email: mizan.rahman@ua.edu \\
  \hfill\break
    \textbf {Alican Sevim, Ph.D. Student}\\
  Department of Civil and Environmental Engineering, FAMU-FSU College of Engineering\\
  2525 Pottsdamer Street, Tallahassee, FL, 32311\\
  Email: as24cq@fsu.edu \\
  \hfill\break
  \textbf{David Bodoh}\\
  Lead Systems Engineer, MITRE Corporation\\
  7525 Colshire Dr, McLean, 22102 Virginia\\
  Email: dbodoh@mitre.org \\
  \hfill\break
  \textbf{Sakib Khan, Ph.D.}\\
  Principal Intelligent Transportation Systems Engineer, MITRE Corporation\\
  7525 Colshire Dr, McLean, 22102 Virginia\\
  Email: sakibkhan@mitre.org \\
  \hfill\break
  \textbf{Li Zhao, Ph.D.}\\
  Department of Civil and Environmental Engineering, University of Nebraska-Lincoln\\
  262K Prem Paul Research Center at Whittier School\\
  2200 Vine Street, Lincoln 68588 Nebraska\\
  Email: lizhao@unl.edu, (402) 472-1928\\
  \hfill\break%
  \textbf{Nathan Huynh, Ph.D.}\\
  Department of Civil and Environmental Engineering, University of Nebraska-Lincoln\\
  262D Prem Paul Research Center at Whittier School\\
  2200 Vine Street, Lincoln 68588 Nebraska\\
  Email: nathan.huynh@unl.edu\\
  \hfill\break%
  \textbf{Eren Erman Ozguven, Ph.D.}\\
  Department of Civil and Environmental Engineering, FAMU-FSU College of Engineering\\
  2525 Pottsdamer Street, Tallahassee, FL, 32311\\
  Email: eozguven@eng.famu.fsu.edu
}

\begin{document}
\maketitle

\section{Abstract}
Proactive safety systems aim to mitigate risks by anticipating potential conflicts between vehicles and enabling early intervention to prevent work zone-related crashes. This study presents an infrastructure-enabled proactive work zone safety warning system that leverages a Digital Twin environment, integrating real-time multi-sensor data, detailed High-Definition (HD) maps, and a historical prediction attention mechanism-based trajectory prediction model. Using a co-simulation environment that combines Simulation of Urban MObility (SUMO) and CAR Learning to Act (CARLA) simulators, along with Lanelet2 HD maps and the Historical Prediction Network (HPNet) model, we demonstrate effective trajectory prediction and early warning generation for vehicle interactions in freeway work zones. To evaluate the accuracy of predicted trajectories, we use two standard metrics: Joint Average Displacement Error (ADE) and Joint Final Displacement Error (FDE). Specifically, the infrastructure-enabled HPNet model demonstrates superior performance on the work-zone datasets generated from the co-simulation environment, achieving a minimum Joint FDE of 0.3228 meters and a minimum Joint ADE of 0.1327 meters, lower than the benchmarks on the Argoverse (minJointFDE: 1.0986 m, minJointADE: 0.7612 m) and Interaction (minJointFDE: 0.8231 m, minJointADE: 0.2548 m) datasets. In addition, our proactive safety warning generation application, utilizing vehicle bounding boxes and probabilistic conflict modeling, demonstrates its capability to issue alerts for potential vehicle conflicts. 

\hfill\break%
\noindent\textit{Keywords}: Proactive Safety, Work Zone Safety, Digital Twin, Trajectory Prediction, HD Maps
\newpage

\section{Introduction}

Roadway safety remains a critical issue, particularly within work zones where vehicle dynamics and driver behavior significantly vary due to altered roadway configurations and proximity to workers. Work zones pose unique risks as drivers navigate narrowed lanes, changed traffic patterns, and changes in speed limits, often in close proximity to roadway workers. Work zone-related crashes have consistently presented higher fatality rates compared to other roadway segments~\cite{FHWASafetyStats}. A 2022 Bureau of Transportation Statistics report shows a 14-year-high crash record of 856 fatal crashes in work zones~\cite{BTS_WorkZoneSafetyData_2025}. A 2022 nationwide study on highway work zone safety by the Associated General Contractors of America (AGC) found that two-thirds of highway contractors have encountered work zone crashes. Additionally, over one-third (37\%) reported project delays as a result, and 7\% experienced worker fatalities. To address these persistent challenges, proactive safety strategies have emerged as a promising solution. A proactive system capable of anticipating vehicle movements can be a potent tool in reducing the likelihood of collisions and enhancing overall safety.


Unlike reactive approaches that respond after an incident, proactive methods leverage real-time data, predictive analytics, and sensor technologies to anticipate and mitigate risks before they escalate. To develop such a proactive safety application, Digital Twin (DT) technologies offer a necessary framework. DT is a real-time synchronized virtual representation of assets and systems present in the real world~\cite{grieves2017digitaltwin}. The flow of information can occur in both directions, from physical to virtual and from virtual to physical, in a DT system. In the context of transportation, such a system is referred to as a Transportation Digital Twin (TDT), where the physical components or processes of the transportation system have a real-time, synchronized, virtual representation. Among the many types of TDT, \textit{Predictive Digital Twin (PDT)} is a class of system that goes beyond representing the real-time state of a physical object; it employs complex modeling to simulate possible future states of the objects~\cite{irfan2024towardstdt, dasgupta2024harnessing}. In this study, we utilize the PDT framework to develop an intelligent, proactive safety system specifically tailored for complex and dynamic work zone scenarios. The goal is not merely to replicate the present state of traffic flow and vehicle behavior, but to anticipate potential crashes before they materialize and enable timely interventions.

This study presents an infrastructure-enabled proactive work zone safety warning system that leverages a PDT environment, integrating real-time multi-sensor data, detailed High-Definition (HD) maps, and a historical prediction attention mechanism-based trajectory prediction model. Our prediction model incorporates the Historical Prediction Attention Network (HPNet)~\cite{tang2024hpnet}, a neural architecture designed to learn from multi-modal trajectory data while accounting for spatio-temporal interactions among vehicles. By capturing both individual vehicle dynamics and inter-vehicle dependencies, HPNet generates future trajectory predictions with high temporal consistency and spatial precision, aligning well with the predictive requirements of a PDT system. Notably, to the best of our knowledge, this study is the first to adapt HPNet for infrastructure-based trajectory prediction, demonstrating that a self-driving car-based trajectory prediction algorithm can be effectively repurposed to operate from the vantage point of roadside infrastructure-mounted sensors. In addition, we developed a proactive safety warning generation application, utilizing vehicle bounding boxes and probabilistic conflict modeling to validate the effectiveness of the PDT framework to issue alerts for potential vehicle conflicts.  For the development and evaluation of the PDT, we leverage a co-simulation platform that tightly integrates Simulation of Urban Mobility (SUMO) with the CAR Learning to Act (CARLA) simulator. This co-simulation framework allows us to synthesize diverse and realistic vehicle trajectory datasets under varied traffic conditions, roadway configurations, and driver behaviors commonly observed in work zones. 



\section{Related Work}
This related work section is divided into two folds: (i) infrastructure-based vehicle trajectory prediction, and (ii) proactive safety warning generation and risk estimation.

\subsection{Infrastructure-based Vehicle Trajectory Prediction}

Trajectory prediction is imperative for many safety critical applications. The premise of the trajectory prediction problem is that, given past motion states of a vehicle, estimate the future positions of that vehicle. These motion states typically include kinematic variables, such as position, velocity, and acceleration. The prediction algorithms may use other relevant information, too, such as historical driving behavior or a detailed map of the road geometry. There are several different approaches to predict vehicle trajectories, which can be broadly classified into data-driven approaches~\cite{chen2025multimodal} utilizing machine learning and deep learning, probabilistic models~\cite{ammoun2009real}, Kalman filters and Bayesian estimators, and physics-based kinematic models~\cite{anderson2021kinematic}. In this study, we refer to the kinematic and probabilistic models as traditional approaches to trajectory prediction. 
Traditional approaches typically overlook inter-vehicle interactions influencing driver behaviors~\cite{lefevre2014survey}. Another important factor that guides the vehicle trajectory is the geometry of the road. The geometry of the road influences drivers as they negotiate roadway curves and turns. Considering the road geometry and how vehicles interact with the road shape and topology provides important cues for the prediction model. A few traditional approaches~\cite{polychronopoulos2007sensor} consider road geometry to provide boundary conditions; however, they lack semantic understanding of the roadway network. 

Early data-driven methods utilized recurrent neural networks (RNNs) like LSTMs to model trajectories~\cite{altche2017lstm}, leveraging the strength of RNNs in modeling temporal dependencies. Temporal convolutional networks (TCNs), an architecture that adapts convolutional neural networks (CNNs) for sequential data and integration with attention mechanism~\cite{vaswani2017attention}, have shown improvements in trajectory prediction~\cite{yuan2023vehicle}. Attention mechanism (AM) in neural networks allows the model to focus on the most relevant parts of the input when processing it, rather than treating all parts equally. The capability to focus on relevant parts of input allows neural networks to achieve ``contextual understanding,'' which led to its use and integration with a wide range of deep learning based models~\cite{huang2022survey}.

In addition to sequential or temporal modeling, recent data-driven deep learning approaches commonly incorporate map information to enhance the understanding of road geometries and provide useful contextual cues. Two different approaches that include map information in trajectory prediction models are: (i) rasterized representations of roadway sections and (ii) detailed vectorized maps that encode lane topology as a graph structure. Early efforts in this area employed CNNs on rasterized images of the roadway, with vehicle trajectories encoded on the image. These methods demonstrated improved prediction accuracy compared to models that rely solely on sequential trajectory data~\cite{nikhil2018convolutional}. More recent approaches leverage high-definition (HD) vector maps~\cite{elghazaly2023high} that provide rich geometric and topological information, such as lane boundaries, connections, and traffic rules. These vector maps are typically encoded as graph structures, where lane segments are encoded as nodes and their connections—including merging, branching, and priority relationships are encoded as edges. Graph structure makes it compelling to apply graph neural networks (GNNs) and graph convolutional networks (GCNs) in trajectory prediction applications. The trajectory prediction model LaneGCN introduced by Liang et. al.~\cite{liang2020learning} is built upon this idea, which showed improved trajectory prediction and provided a foundation for subsequent studies. Studies, such as VectorNet~\cite{gao2020vectornet}, QCNet~\cite{zhou2023query}, and HiVT~\cite{zhou2022hivt}, have demonstrated significant advancements in trajectory prediction modeling by leveraging HD vector maps, attention mechanisms, and graph convolutional networks. In addition, increased industry partnership and benchmark datasets, such as Waymo~\cite{waymo_motion_prediction_challenge_2024}, Argoverse~\cite{Argoverse2}, INTERACTION~\cite{2019-interactiondataset}, accelerated the development of increasingly better performing trajectory prediction models and a mechanism to compare the results of different models. 

Most benchmark datasets primarily target self-driving vehicles, which has to anticipate movements of surrounding vehicles for its path planning module and safety. This incentivizes the development of sophisticated trajectory prediction models, which are predominantly evaluated in self-driving contexts. Recently, HPNet~\cite{tang2024hpnet} ranked first in the interaction dataset due to its novel triple factorized attention and utilization of historical predictions for better temporal consistency. However, the advances in prediction model architectures are also applicable to other domains, particularly roadside infrastructure-based trajectory prediction for safety applications~\cite{zhao2025vehicle}. A key strength of HPNet's multi-modal output is its ability to capture the inherent variability in human driving behavior. Human drivers exercise free will, often making subtle, discretionary choices that result in slightly different yet equally plausible trajectories. For instance, a driver may choose to navigate a work zone by slightly adjusting their path or speed, with each choice having comparable likelihood. HPNet's multi-modal predictions reflect this by generating multiple trajectory hypotheses, each adhering to physical constraints and map context, such as lane boundaries and temporary signage. These trajectories, while distinct, exhibit high agreement due to the constrained nature of work zone environments, where variability is subtle and bounded by roadway geometry and traffic rules. This balance allows HPNet to model the nuanced decision-making of human drivers while maintaining temporal coherence and spatial consistency across predictions. The primary contribution of this study is to demonstrate the feasibility and effectiveness of adapting HPNet for infrastructure-based trajectory prediction in a highway work zone setting.


\subsection{Proactive Warning Generation and Risk Estimation}

Accurate estimation of the risk of vehicle conflict and translating it into effective warnings remains a central challenge in proactive transportation safety systems. To this end, recent research has increasingly adopted probabilistic modeling frameworks to address the uncertainties inherent in vehicle trajectories, driver behaviors, and dynamic traffic contexts such as highway work zones. A foundational principle in modern risk estimation is trajectory-level probabilistic modeling. Gaussian processes and Hidden Markov Models have been combined to predict vehicle motion under uncertain conditions to produce collision risk estimations that generalize beyond fixed assumptions of linearity or constant speed~\cite{tay2012probabilistic}. Similar probabilistic approaches have been applied in pedestrian-vehicle contexts, where Gaussian process regression and maneuver likelihoods allow for granular conflict quantification with lower computational costs than deep learning models~\cite{li2023probabilistic}. These frameworks outperform traditional surrogate safety measures (SSMs), such as time-to-collision (TTC), which often rely on rigid motion assumptions. 

Although PET (Post Encroachment Time) and TTC remain fundamental to surrogate safety measure (SSM) approaches, many researchers have enhanced their predictive capability by integrating these indicators within probabilistic modeling frameworks that better capture traffic’s inherent randomness. Among these, Extreme Value Theory (EVT) has emerged as a powerful tool for quantifying rare but critical safety events~\cite{azevedo2015using, jiao2025unified, nazir2024effects, niu2024cross, songchitruksa2006extreme, tarko2021unifying, zheng2014freeway}. \citep{alozi2022evaluating} analyzed large-scale AV-pedestrian interaction datasets from platforms like Motional and Lyft, using PET and TTC to derive risk profiles. Their findings indicated elevated risk exposure in the early stages of autonomous vehicle deployment, with an estimated 4–5.5 collisions per million vehicle kilometers traveled (VKT). When contextual variables were added, this range narrowed to 2.3–3.7 per million VKT, offering greater statistical confidence. Their results underscore how EVT-based probabilistic models can accommodate behavioral variability and provide scalable risk metrics in rapidly evolving transportation systems. A recent framework integrates backward and stochastic forward reachable sets to distinguish guaranteed-safe states from probabilistically risky ones with the intent of enhancing both safety guarantees and precision in dynamic conflict identification~\cite{wang2024reachability}. Such hybrid techniques help mitigate over-conservatism in earlier reachability analyses, while supporting confidence-aware risk predictions essential for real-world deployment. On the other hand, \citep{jiao2025unified} developed a probabilistic model that characterizes traffic conflicts as context-sensitive extremes within road user behavior. Unlike conventional SSM approaches that apply static thresholds based on conflict type, their method reframes conflict risk estimation as a statistical learning problem, where extreme value inference enables more adaptive detection. In this study, to develop proactive safety application, we leverage a probabilistic model that characterizes traffic conflicts and present a mathematical framework for predicting pairwise vehicle conflicts based on multi-step trajectory forecasts.

\section{Predictive Digital Twin Framework}
\begin{figure}[H]
    \centering
    \includegraphics[width=0.85\linewidth]{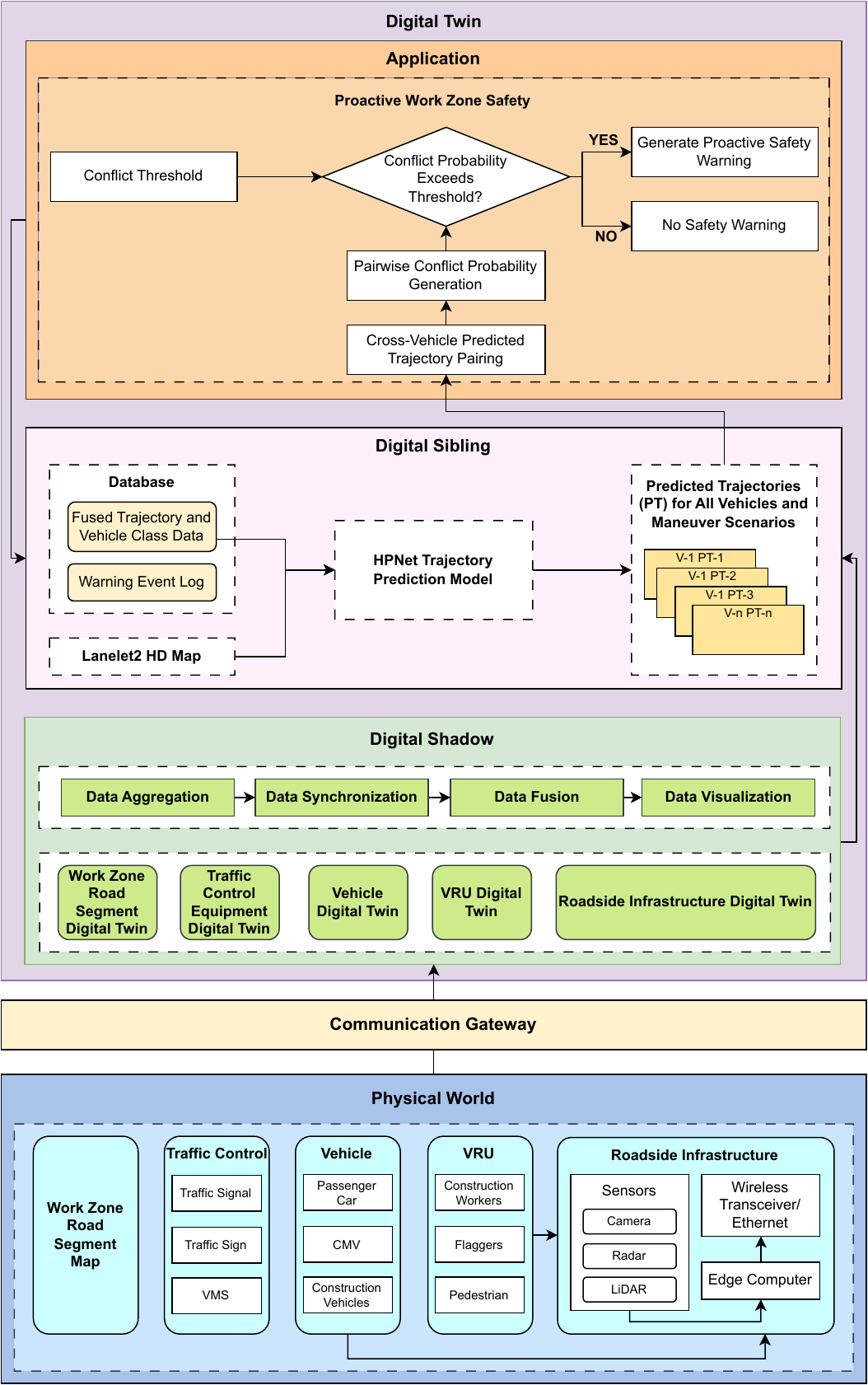}
    \caption{Predictive Digital Twin (PDT) framework for proactive work zone safety}
    \label{fig:PDT}
\end{figure}

Figure~\ref{fig:PDT} presents a predictive digital twin (PDT) framework for proactive work zone safety using roadside sensors data, which comprises three layers: 1) physical world; 2) DT; and 3) communication gateway. The physical world serves as the bottom layer and encompasses all transportation-related physical objects, such as vehicles, work zone road segments, workers, and roadside sensors. The top layer represents the DT, which consists of digital shadow, digital sibling, and real-time proactive safety applications. The top and bottom are connected by the communication gateway, facilitating real-time data exchange between these two layers. Roadside sensors can be equipped with computing devices and sensors (e.g., cameras, RADAR, and LiDAR) for precise detection and continuous trajectory generation of vehicles and workers within the work zone roadway segment. DT layer consists of digital shadow, digital sibling and proactive safety applications. In the context of PDT, digital shadow is the digital replica of vehicles and workers trajectories. Digital shadow can be created and updated in real-time by aggregating, synchronizing, and fusing data transmitted from the physical world. Digital shadow receives sensor data from the physical world and stores processed data in the digital sibling database, but it cannot transmit information back to the physical world.
Digital sibling has three basic components: 1) Database; 2) Prediction model; 3) Work zone scenario with roadway geometry and regulatory elements encoded in a Lanlet2 HD Map. The prediction model has the capability of predicting all plausible vehicle trajectories with different probabilities.  Finally, the proactive safety application module calculates conflict probability for all combinations of predicted trajectory pairs and uses a conflict decision threshold for proactive safety alert generation. The following subsections detail the development pipeline, prediction model, and proactive safety application.

\subsection{Integration of Simulation, Prediction, and Warning generation}
\begin{figure} [H]
    \centering
    \includegraphics[width=0.67\linewidth]{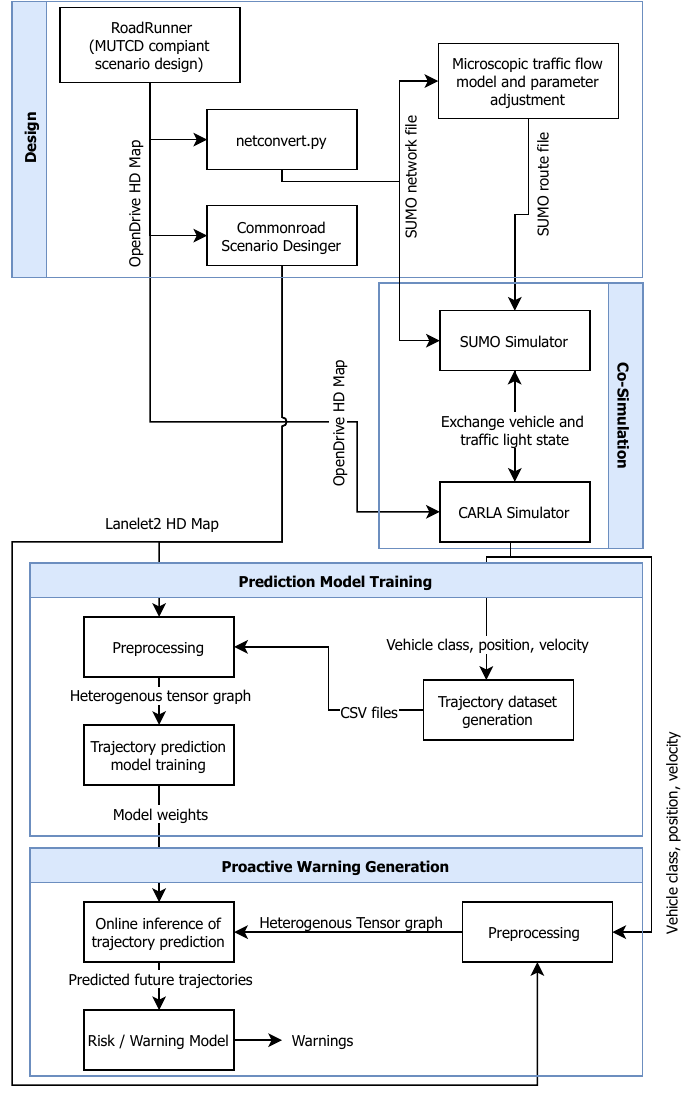}
    \caption{Framework integrating simulation, prediction, and warning generation under a PDT framework}
    \label{fig:traj-predict-sim-flow}
\end{figure}

This study presents a comprehensive pipeline for developing a proactive warning system for work zone environments within a predictive digital twin framework, integrating scenario design, co-simulation, prediction model training, and warning generation, as illustrated in Figure~\ref{fig:traj-predict-sim-flow}. The process begins with the Design stage, where road networks and traffic scenarios are created using tools like RoadRunner and Commonroad Scenario Designer, generating OpenDrive and Lanelet2 HD maps, alongside SUMO network and route files, with microscopic traffic flow models tailored for realistic simulations. In the Co-Simulation stage, the SUMO and CARLA simulators interact to exchange vehicle and traffic light states, creating a realistic environment for data collection and testing that mimics real-world implementation processes. The Prediction Model Training stage involves generating trajectory datasets from simulations, preprocessing them, and training a trajectory prediction model. Finally, the Proactive Warning Generation stage utilizes the trained model for online inference, enabling real-time deployment by predicting future trajectories and generating proactive warnings using vehicle data.

\subsection{Trajectory Prediction of All Plausible Future Maneuverings}

In this study, we utilized HPNet, developed by Tang et al.~\cite{tang2024hpnet}. The key contribution of the authors is the development of temporally consistent trajectory prediction of all plausible future maneuvers. Many conventional trajectory prediction models treat each future timestep independently, leading to temporally incoherent predictions where the sequence of predicted positions lacks smoothness or dynamic feasibility. Such frame-wise decoupling results in discontinuities in velocity and heading, particularly under multi-modal settings where multiple hypotheses are generated. HPNet mitigates this issue through a hierarchical decoder that models future trajectories as temporally dependent sequences. Let the observed trajectory of an vehicle be denoted as \( \mathbf{X}_{1:T_{\text{obs}}} = \{\mathbf{x}_1, \mathbf{x}_2, \ldots, \mathbf{x}_{T_{\text{obs}}} \} \), where \( T_{\text{obs}} \) is the number of historical timesteps, and \( \mathbf{x}_t \in \mathbb{R}^2 \) represents the 2D position at time \( t \). The goal is to predict \( K \) plausible future trajectories \( \{ \hat{\mathbf{Y}}^{(k)} \}_{k=1}^K \), where each \( \hat{\mathbf{Y}}^{(k)} = \{ \hat{\mathbf{y}}^{(k)}_1, \hat{\mathbf{y}}^{(k)}_2, \ldots, \hat{\mathbf{y}}^{(k)}_{T_{\text{pred}}} \} \) is a temporally ordered sequence of future positions. Specifically, the model autoregressively decodes future positions, where the prediction at timestep $t+1$ is conditioned not only on the historical context $\mathbf{X}_{1:T_\text{obs}}$ but also on the predictions up to timestep $t$, i.e., $\hat{\mathbf{Y}}_{1:t}$. This design enforces internal temporal consistency and ensures that trajectory evolution follows physically plausible motion patterns. Moreover, HPNet's multi-head attention modules, such as historical and mode attention, allow the model to selectively focus on salient temporal cues and disambiguate agent intent across the full prediction horizon.

The HPNet model was originally developed for an ego-centric self-driving vehicle, where it needs to assess the future trajectories of surrounding vehicles. Unlike its initial application with onboard sensors, our implementation adapted HPNet for roadside infrastructure equipped with stationary LiDAR,radar and/or camera sensors in a work zone scenario. HPNet employs a hierarchical, spatial-temporal deep learning framework to extract and model complex interactions and movement patterns. At its core, the network integrates Graph Neural Networks (GNNs), specifically utilizing Graph Attention Networks (GATs), to encode spatial and temporal interactions between vehicles and the roadway section. It processes the HD Map as a vector graph with attention modules. This graph-based representation enables HPNet to dynamically emphasize relevant spatial relationships and temporal dependencies, improving the predictive performance of the model. Multiple attention layers were employed to dynamically weigh various spatiotemporal features, thereby facilitating multimodal trajectory predictions that accommodate inherent uncertainties in future vehicle movements.

Attention mechanism~\cite{vaswani2017attention} is a core component of the model. Attention mechanisms enable the model to selectively focus on relevant parts of the input data when generating predictions, significantly enhancing predictive accuracy and contextual awareness. The attention mechanism operates on three fundamental parameters: Query (Q), Key (K), and Value (V). Their roles are as follows: Query (Q) represents the elements seeking information. Key (K) represents the actual informational content provided in response to the queries, weighted by the relevance determined through query-key matching. The input information are tokenized by the embedding process to be used as K, Q, and V matrices. The model inputs include trajectories of the vehicles (sequence of position, heading, and velocity) encoded by an Multi Layer Perceptron (MLP) layer into location-dependent vehicle feature embeddings (\texttt{E\_a}). Additionally, static attributes, such as vehicle length, width, and type, are incorporated to enrich the representation. Map attributes, including lane boundaries and center lines, are encoded into map feature embeddings (\texttt{E\_m}). Relative features, capturing spatial relationships, such as distances, angles, and heading, and differences between agents and map elements, are encoded into relative feature embeddings (\texttt{E\_e}).

\begin{figure}[!ht]
    \centering
    \includegraphics[width=\linewidth]{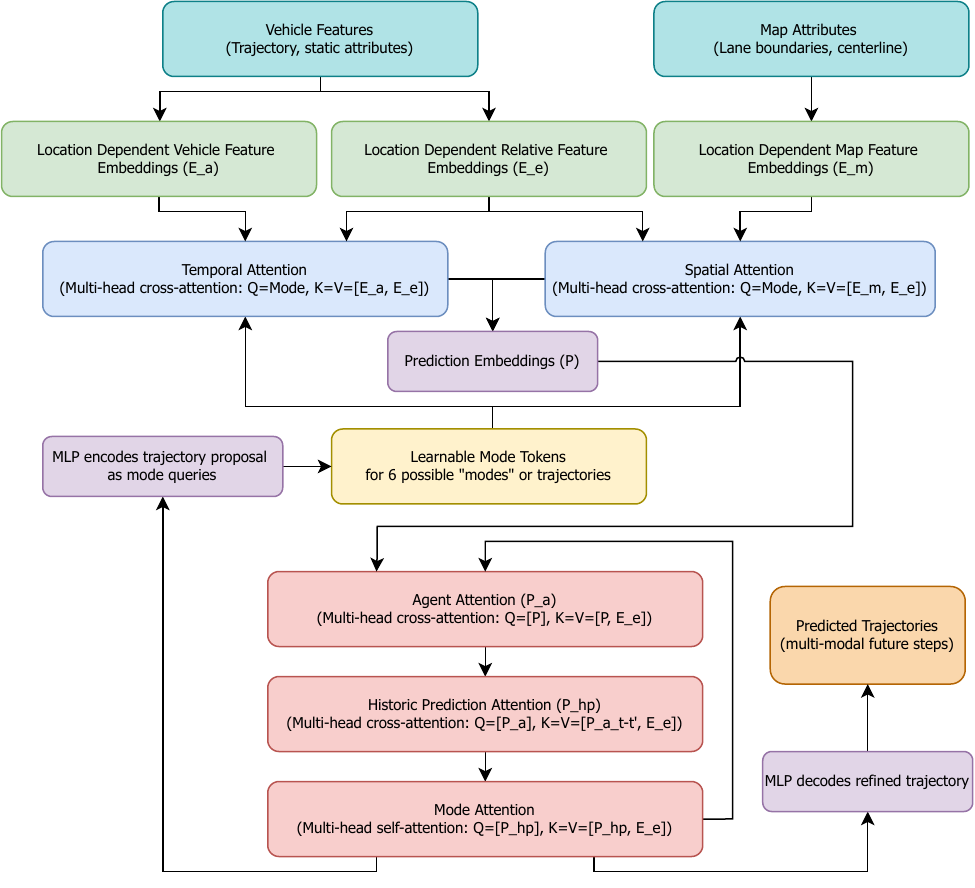}
    \caption{Historical trajectory prediction model architecture with multi-stage attention for temporally consistent trajectory prediction}
    \label{fig:hpnet-architecture}
\end{figure}

The model architecture, as depicted in Figure~\ref{fig:hpnet-architecture}, processes the inputs through a series of attention-based layers to generate and refine multi-modal trajectory predictions. The process begins with learnable mode tokens, representing \( K \) possible future trajectories for each agent, which act as queries in attention mechanisms. Temporal attention uses multi-head cross-attention, with mode tokens as queries and vehicle/relative embeddings (\texttt{E\_a}, \texttt{E\_e}) as keys and values, to capture historical vehicle dynamics over time. Spatial attention employs cross-attention with mode tokens as queries and map/relative embeddings (\texttt{E\_m}, \texttt{E\_e}) to integrate roadway context, aligning predictions with lane constraints. Mode attention further refines predictions through three sub-layers: agent attention (P\_a), which models inter-agent interactions using self-attention on prediction embeddings (\texttt{P}) and relative features; historic prediction attention (P\_hp), which ensures temporal coherence by capturing dependencies within predicted trajectories; and mode attention, which disambiguate multiple plausible trajectory hypotheses by modeling interactions across modes.

The trajectory prediction process unfolds in two stages: proposal generation and refinement. In the proposal stage, an MLP (\texttt{traj\_propose}) decodes mode embeddings (\texttt{m\_embs}) into initial multi-modal trajectory proposals (\texttt{traj\_propose}) in global coordinates, with shape \( [(N_1, \ldots, N_b), H, K, F, 2] \), where \( N_1, \ldots, N_b \) denote agents per batch, \( H \) is historical timesteps, \( K \) is modes, \( F \) is future timesteps, and 2 represents 2D coordinates. These proposals define anchor points by selecting the middle future timestep (\( F//2 \)) to compute anchor positions (\texttt{n\_position}) and headings (\texttt{n\_heading}), which are transformed into local coordinates and encoded into anchor embeddings (\texttt{n\_embs}) using an MLP (\texttt{proposal\_to\_anchor}). In the refinement stage, anchor embeddings undergo another temporal attention stage to integrate historical agent data, a spatial attention to incorporate map context, and mode attention to refine interactions across agents, historical steps, and modes. An MLP (\texttt{traj\_refine}) generates a correction term (\texttt{traj\_refine}), which is added to the initial trajectory proposal and transformed back to global coordinates using \texttt{transform\_traj\_to\_global\_coordinate}, yielding the final output (\texttt{traj\_output}). This two-stage approach ensures contextually informed and temporally consistent predictions, overcoming the limitations of frame-wise decoupled models.

\subsection{Proactive Safety Warning Generation}
\label{sec:warning-generation-method}
In this section, we present a mathematical framework for predicting pairwise vehicle conflicts based on multi-step trajectory forecasts. Each vehicle is represented geometrically as a rigid rectangular bounding box situated in a two-dimensional Euclidean space. The center of the vehicle is denoted by a time-indexed sequence of predicted positions $(x_i^{(k)}, y_i^{(k)})$, accompanied by an orientation angle $\psi_i^{(k)}$, where $k$ indicates the prediction step and $i$ indexes the vehicle within the set of vehicles present in a given time frame. The shape of each vehicle is defined by fixed parameters for length $l$ and width $w$, and the corresponding bounding box is constructed by applying a planar rotation matrix to a predefined set of local coordinates to represent the corners and midpoints of the vehicle’s physical envelope. The transformation into the global coordinate system is performed by rotating these points using $\psi_i^{(k)}$ and translating them to the global position $(x_i^{(k)}, y_i^{(k)})$. This transformation for a point $(u,v)$ in the vehicle’s local coordinate frame is expressed as $(x_i^{(k)}, y_i^{(k)}) + \mathcal{R}(\psi_i^{(k)}) \cdot (u, v)^T$ where $\mathcal{R}(\psi)$ denotes the standard two-dimensional rotation matrix. The notation $(u,v)^T$ denotes the column vector form of a point defined in the vehicle's local coordinate frame. While $(u,v)$ represents a 2D point as a row vector, its transpose $(u,v)^T = \begin{bmatrix} u \\ v \end{bmatrix}$ is required for matrix multiplication with the 2x2 rotation matrix $\mathcal{R}(\psi)$. This transformation rotates the local point according to the vehicle’s orientation $\psi$ to be accurately mapped into global coordinates:

\begin{linenomath}
\begin{flalign}
&\mathcal{R}(\psi_i^{(k)}) \cdot \begin{bmatrix} u \\ v \end{bmatrix}
= \begin{bmatrix}
\cos\psi & -\sin\psi \\
\sin\psi & \cos\psi
\end{bmatrix} \cdot \begin{bmatrix} u \\ v \end{bmatrix} 
=  \begin{bmatrix} u\cos\psi - v\sin\psi \\ u\sin\psi + v\cos\psi \end{bmatrix} &
\end{flalign}
\end{linenomath}

For each discrete time frame and prediction step $k$, we evaluate all unique unordered pairs of vehicles $(v_i, v_j)$, with $i < j$, to estimate the likelihood of a conflict. The minimum Euclidean distance between any pair of transformed bounding box points associated with $v_i$ and $v_j$ is computed and denoted as $d_{ij}^{(k)}$:

\begin{linenomath}
\begin{flalign}
&d_{ij}^{(k)} = \min_{a,b} \left\| p_{ia}^{(k)} - p_{jb}^{(k)} \right\|_2 &
\end{flalign}
\end{linenomath}
\noindent
where $p_{ia}^{(k)}$ and $p_{jb}^{(k)}$ are the transformed global coordinates of the $a$-th and $b$-th bounding box points for vehicles $i$ and $j$, respectively. In a conflict hierarchy, the closer the proximity between approaching road users, the less safe they are and the higher the probability of a potential collision~\cite{jiao2024unified}. Hence, the scalar distance $d_{ij}^{(k)}$ is subsequently transformed into a probabilistic conflict score via an exponential decay function of the form:
\begin{linenomath}
\begin{flalign}
&P_{ij}^{(k)} = \exp\left(-\frac{d_{ij}^{(k)}}{\lambda} \right) &
\end{flalign}
\end{linenomath}
\noindent
where $\lambda > 0$ is a decay constant governing the rate at which conflict likelihood diminishes with spatial separation. This function handles the conflict probability approaches unity as the separation approaches zero and decays asymptotically toward zero as the distance increases, and at the same time provides a smooth, differentiable risk measure based on physical proximity. A pairwise interaction is categorized as a potential conflict if the distance $ d_{ij}^{(k)}$ falls below a designated spatial threshold $\delta_d$. In addition, a secondary criterion identifies high-risk interactions based on the conflict probability score, wherein an interaction is flagged if $P_{ij}^{(k)} > \delta_P$ for some chosen probability threshold  $\delta_P \in (0,1)$. These criteria enable the separation of low-risk and high-risk conflict candidates, facilitating downstream filtering for scenario analysis or adjusting mechanisms. The computational process iterates over all time frames and all prediction horizons by maintaining a structured record of each evaluated vehicle pair, their geometric separation, the associated conflict probability, and whether the pair meets the criteria for conflict or high-risk designation. It has directional sensitivity in conflict modeling. For example, vehicles approaching each other head-on or at acute angles may present higher conflict risk despite moderate centroidal distances. The alignment of bounding box surfaces implicitly captures such configurations for a more accurate detection of spatial convergence zones. This is especially critical in environments with complex maneuvers, such as work zones, intersections, or lane merges, where lateral and angular interactions are dominant. From a mathematical standpoint, bounding-box distances generalize center-based metrics by reducing to the latter in the degenerate case where vehicle dimensions vanish. As such, they preserve the analytical simplicity of Euclidean norms while enriching the geometric expressiveness of the model. 


\section{SUMO-CARLA Experimental Setup}
In this study, we utilize SUMO and CARLA simulators, two widely accepted open-source simulators for traffic, vehicle dynamics, and sensor simulation. SUMO provides realistic large-scale traffic simulation capabilities with standard microscopic traffic flow models and is particularly adept at simulating detailed merging behaviors in constrained environments like work zones \cite{lopez2018microscopic}. In this study, CARLA complements SUMO by offering high-fidelity sensor simulation, such as lidar, radar, and camera. We configure SUMO with merging points. SUMO's car-following and lane-changing models are adjusted to match observed real-world driver behaviors within work zones closely. These behaviors include realistic acceleration, deceleration, and merging maneuvers validated by traffic flow studies \cite{kesting2008adaptive}.

SUMO’s default lane-changing model (LC2013) executes instantaneous lane changes, insufficient for capturing realistic merging dynamics near work zone lane closures. We address this by employing SUMO’s sublane model (SL2015), enabling continuous, gradual lateral movements rather than abrupt lane changes. In the sublane model, the road is conceptually divided into many fine-grained lateral subdivisions (sublanes), allowing vehicles to occupy fractional positions within a lane or between lanes. This enables the modeling of lane changes as a time-dependent process influenced by vehicle dynamics, rather than as a discrete event. To activate the sublane model, SUMO requires to be configured with the \texttt{--lateral-resolution <FLOAT>} parameter set to a value smaller than the lane width, which is 0.25 meters in our case. With sublane model enabled, SUMO switches the lane changing model from the default LC2013 to SL2015 model, which can use the lane subdivisions for smoother and realistic transitions. Once enabled, the vehicles can initiate a lane change by steering toward the target lane, and the movement unfolds over multiple simulation steps, as a gradual process. In addition to LC2013 model parameters, the trajectory and duration of this lateral transition are governed by additional vehicle parameters, such as lcSublane, lcAssertive, lcImpatience, lcTimeToImpatience, lcAccelLat, latAlignment, and lcMaxSpeedLatStanding. These parameters together shape how vehicles behave during lane changes. For instance, lcSublane controls the vehicle’s preference for staying aligned laterally within a lane, while lcAssertive and lcPushy determine how willing a vehicle is to accept tighter gaps or initiate lane changes in partially blocked spaces. lcImpatience dynamically scales the assertive and pushy tendencies as time progresses, with lcTimeToImpatience setting how long it takes to reach peak impatience. lcAccelLat restricts the lateral acceleration to reflect realistic dynamics, and latAlignment allows vehicles to aim for specific offsets within the lane (e.g., center, left, or right). Lastly, the parameters lcMaxSpeedLatStanding and lcMaxSpeedLatFactor control the lateral speed envelope, scaling lateral velocity based on vehicle speed to prevent unrealistic side movements. Together, these parameters enable nuanced, gradual, and context-aware lateral maneuvers that are critical for simulating realistic merging behavior in work zones where vehicles should respond to upstream lane closures. This modeling enhancement is especially useful because it allows us to accurately capture the interactions between merging vehicles and through traffic, as well as model behaviors such as hesitation, courtesy yielding, and multi-vehicle negotiations near the taper zone of the work zone. Additionally, the continuous, dynamics and time-evolving lateral motion representation allows for better integration with trajectory prediction algorithms, which requires smooth and physically plausible vehicle motions.

\begin{figure}[!ht]
    \centering
    \includegraphics[width=0.9\linewidth]{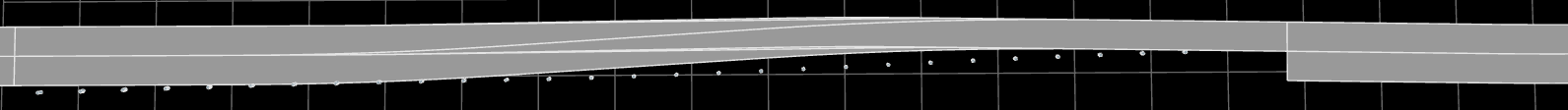}
    \caption{Work zone scenario with lane closure and taper designed in RoadRunner}
    \label{fig:map}
\end{figure}

\begin{figure}[!ht]
    \centering
    \includegraphics[width=0.5\linewidth]{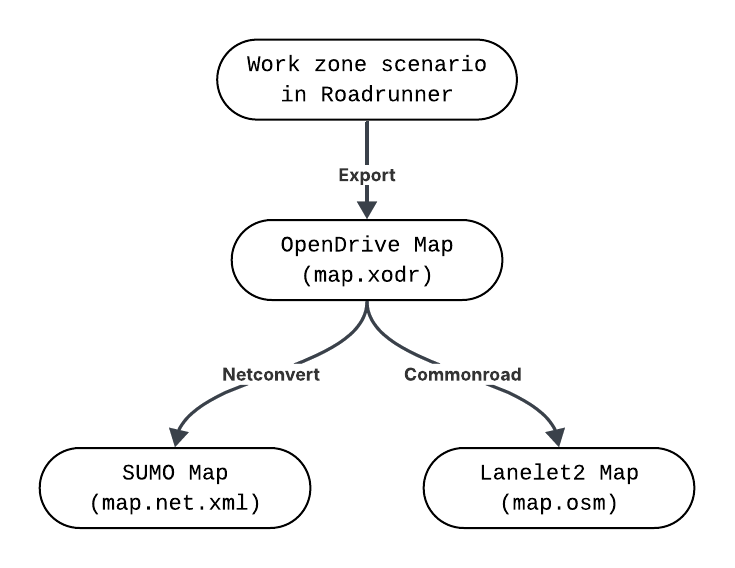}
    \caption{Map conversion workflow from OpenDRIVE to SUMO and Lanelet2 formats}
    \label{fig:map-conversion}
\end{figure}

The road network for SUMO-CARLA co-simulation is created using Mathworks Roadrunner software, following Manual on Uniform Traffic Control Devices (MUTCD) guidelines for lane closure in a work zone including speed reduction and taper lengths~\cite{brown2000mutcd}. Figure~\ref{fig:map} shows a portion of the work zone map created in Roadrunner. The map geometry can be exported to OpenDrive format (map.xodr) HD Map, which is a direct input to the CARLA simulator. CARLA supports native parsing of the OpenDrive format road networks. CARLA uses the file and constructs corresponding 3D meshes for the road topology and renders in 3D simulation environment. However, the SUMO simulator cannot take an OpenDrive map as input. It uses its own XML format for network definitions. On the other hand, the trajectory prediction module uses yet another HD Map format, Lanelet2. The figure~\ref{fig:map-conversion} illustrates how the map has been converted to be used by SUMO, CARLA, and the trajectory prediction module. 

To facilitate SUMO-CARLA co-simulation, SUMO's \texttt{netconvert} tool was used for converting OpenDrive maps into SUMO-compatible networks. \texttt{Netconvert} parses the road geometry and attributes from the XODR format and generates an equivalent network topology interpretable by SUMO. The `net.xml' file serves as the road network for SUMO, enabling us to simulate traffic and synchronize it with CARLA through our co-simulation bridge. This allows us to simulate traffic interactions, including vehicle behavior through the work zone in SUMO, while maintaining physical realism in the CARLA environment, so that it can be perceived by sensors such as LiDAR, radar, and camera.

The adopted HPNet model has a module that parses the Lanelet2 format HD Map to create its internal vector graph representation of the map. We converted the OpenDrive format map into Lanelet2 using the open-source tool Commonroad Scenario Designer~\cite{maierhofer2021commonroad}. Commonroad provides a common abstraction for elements commonly found in different HD Map standards, which in turn enables seamless conversion between standard map formats.

\section{Data Generation}
Vehicle trajectory data were captured using stationary LiDAR, radar and camera sensors strategically positioned along the roadside near active construction zones. The fusion of all three sensors provides vehicle classification and high-resolution three-dimensional positional data (X, Y, Z coordinates) of passing vehicles. All three sensor modalities collect data at a frequency of 10 Hz, which facilitates high temporal resolution and continuous vehicle tracking. In our simulation setup, we also collect the ground-truth data of all the vehicles at 10 Hz (0.1-second time interval), which is used for training the HPNet model. 

\begin{table}[H]
    \caption{Description of Columns in the \texttt{trajectory\_data\_case\_xxx.csv} File}
    \label{tab:vehicle_tracks}
    \centering
    \begin{tabular}{llp{10cm}}
    \hline
    \textbf{Column} & \textbf{Attribute} & \textbf{Description} \\
    \hline
    1 & \texttt{track\_id} & The ID associated with each vehicle in the scene. The \texttt{track\_id} ranges from 1 to \textit{n}, where \textit{n} is the number of vehicles in the scene. \\
    2 & \texttt{timestamp\_ms} & The timestamp associated with the trajectory, ranging from 100 ms to at most 4000 ms, conforming to the 4-second trajectory recording. \\
    3 & \texttt{frame\_id} & For each vehicle (\texttt{track\_id}), the number of frames the vehicle appeared in the data. It ranges from 1 to at most 40, as a vehicle appears at most 40 times within the 4-second period recorded at 100 ms intervals. \\
    4 & \texttt{agent\_type} & The category of the vehicle (e.g., car, truck). \\
    5 & \texttt{x} & The \textit{x} position of the vehicle (meters) in the local coordinate system. \\
    6 & \texttt{y} & The \textit{y} position of the vehicle (meters) in the local coordinate system. \\
    7 & \texttt{vx} & The \textit{x} component of the velocity (meters per second) in the local coordinate system. \\
    8 & \texttt{vy} & The \textit{y} component of the velocity (meters per second) in the local coordinate system. \\
    9 & \texttt{psi\_rad} & The yaw angle heading of the vehicle (radians). \\
    10 & \texttt{length} & The length of the vehicle in meters. \\
    11 & \texttt{width} & The width of the vehicle in meters. \\
    \hline
    \end{tabular}
\end{table}

The structure of INTERACTION-dataset~\cite{2019-interactiondataset} was followed as a reference when building our dataset using the simulation data. The dataset consists of comma-separated-value (CSV) files, each file containing a 4-second-long scenario. Since the data is recorded at 10 Hz, and there is 18-22 vehicles at any given time in the simulation, each CSV file has approximately 800 rows of data. A total of 20,000 seconds of driving simulation data was generated, which results in 5,000 individual cases, each stored as a separate CSV file. These cases collectively represent approximately 5.56 hours of continuous simulated driving. The CSV files are in the indexed with a `case\_id` (e.g., trajectory\_data\_case\_10.csv). The description of the data fields are described in Table~\ref{tab:vehicle_tracks}.

\section{Evaluation Outcomes}

\subsection{Evaluation Metrics}

To evaluate the accuracy of predicted trajectories, we use two standard metrics widely used in the literature: Average Displacement Error (ADE) and Final Displacement Error (FDE). Let $\mathbf{y}_t = (x_t, y_t)$ denote the ground-truth position at time step $t$, and let $\hat{\mathbf{y}}_t = (\hat{x}_t, \hat{y}_t)$ represent the predicted position at the same time step. Given a prediction horizon of $T$ time steps, the metrics are defined as follows:

\begin{linenomath}
  \begin{flalign}
    &\mathrm{ADE} = \frac{1}{T} \sum_{t=1}^{T} \left\| \hat{\mathbf{y}}_t - \mathbf{y}_t \right\|_2 &\\
    &\mathrm{FDE} = \left\| \hat{\mathbf{y}}_T - \mathbf{y}_T \right\|_2 &\\ 
  \end{flalign}
\end{linenomath}

\noindent
The ADE measures the average Euclidean distance (L2-norm distance) between the predicted and ground-truth positions across the entire prediction horizon, providing a holistic view of the model's performance over time. In contrast, the FDE focuses solely on the Euclidean distance at the final prediction time step, capturing the model’s accuracy in estimating the vehicle's ultimate destination. These metrics jointly capture both spatial consistency and endpoint-oriented accuracy. A low value of ADE and FDE is desired for a trajectory prediction model. 

\subsection{Training and Validation of Trajectory Prediction Model}
\begin{figure}[!ht]
    \centering
    \includegraphics[width=0.8\linewidth]{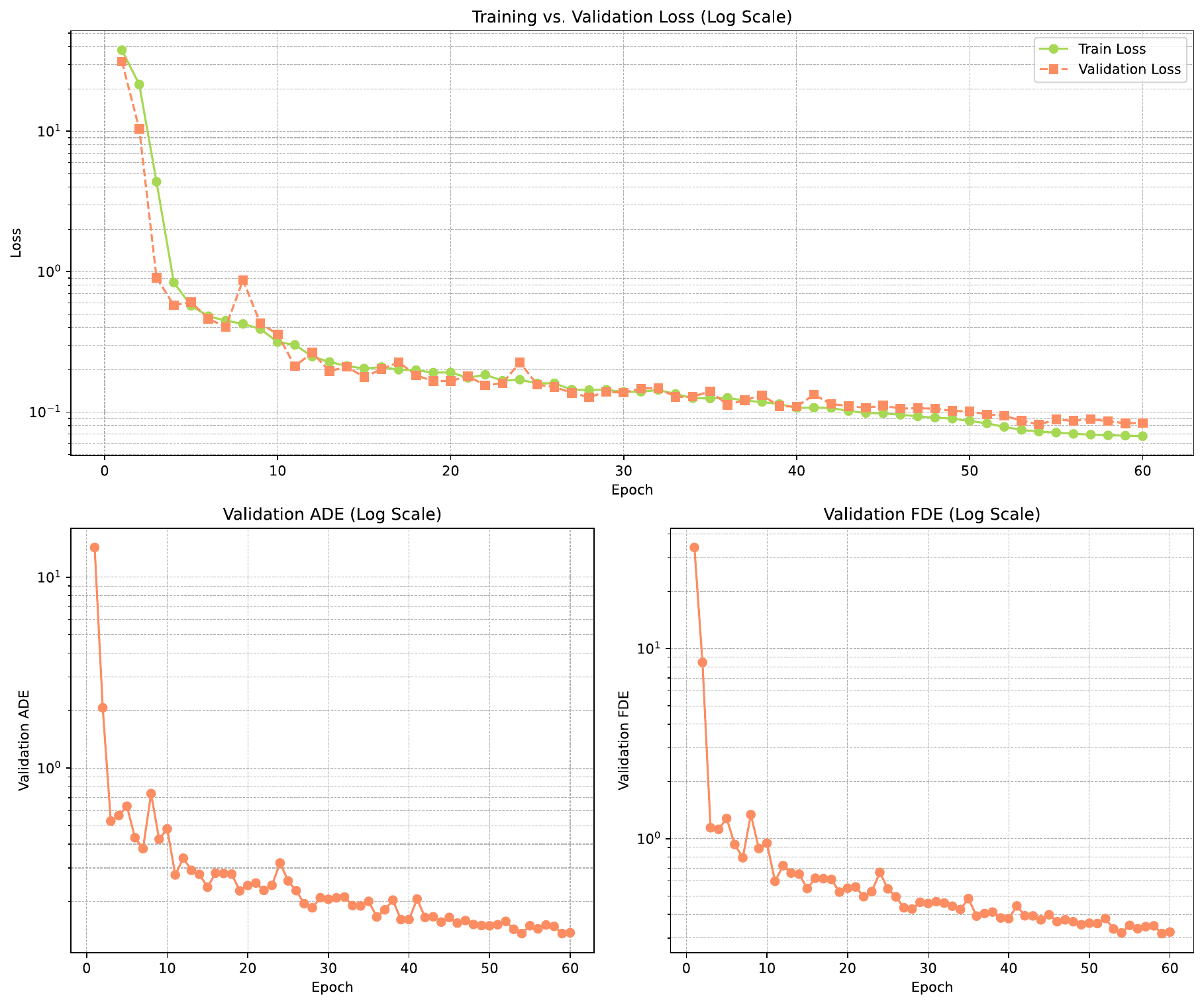}
    \caption{Training and validation loss with ADE/FDE trends across epochs}
    \label{fig:training-graph}
\end{figure}

We trained the HPNet model for 60 epochs using the Adam optimizer with an initial learning rate of 5e-4, decayed by a factor of 1e-4 after 4 warmup epochs. As shown in Figure~\ref{fig:training-graph}, the training loss drops sharply during the first 10 epochs—from over 37 down to approximately 0.39—before decaying more gradually to a final value of 0.06. The validation loss follows a very similar trajectory, decreasing from an initial 31 to 0.08 by epoch 60. The close alignment of the two curves, especially beyond epoch 15, indicates that the model converges reliably without significant overfitting. The vertical axis of the figure is in logarithmic scale to capture the rapid initial convergence and the later fine-scale improvements.  

We further evaluate predictive accuracy using two standard trajectory‐prediction metrics:ADE and FDE. Figure~\ref{fig:training-graph} also illustrates that the minJointADE decreases from roughly 14.0 m at epoch 1 to 0.13 m at epoch 60, while the minJointFDE falls from about 32 m down to 0.322 m over the same training horizon. Both metrics stabilize after epoch 30, confirming that the model’s improvements plateau once the network has sufficiently learned the underlying motion patterns in the dataset.

Overall, the parallel decline in training and validation losses—combined with the monotonic reduction in ADE and FDE—demonstrates that HPNet effectively captures vehicle motion characteristics. The absence of a widening gap between training and validation losses, along with the smooth decay of displacement errors, suggests generalization and supports the choice of our training schedule and its associated hyperparameters.

JointADE is the measure of the average L2-norm distance across all vehicles' predicted and actual trajectories, while minJointADE indicates the closest mode (of 6) to the ground truth. On the other hand, MinJointFDE measures the average L2-norm distance at the final time step for all vehicles. These metrics are widely adopted in benchmarking trajectory prediction performance. Table~\ref{tab:hpnet-performance} compares these metrics from our infrastructure-based work zone simulation dataset with the Argoverse and Interaction datasets. The adapted HPNet model demonstrates superior performance on the work-zone simulation dataset, achieving a minJointFDE of 0.3228 meters and a minJointADE of 0.1327 meters, significantly lower than the benchmarks on the Argoverse (minJointFDE: 1.0986 m, minJointADE: 0.7612 m) and Interaction (minJointFDE: 0.8231 m, minJointADE: 0.2548 m) datasets. This performance disparity can be attributed to the varying complexity of the datasets. The benchmark datasets, Argoverse and Interaction, are derived from real-world data, capturing the diverse and unpredictable behavior of human drivers. In contrast, the work zone simulation dataset, while instrumental in developing and testing the entire pipeline and designed to be as realistic as possible, relies on an underlying microscopic traffic flow model. This model inherently lacks the full spectrum of variability exhibited by real-world human drivers, potentially leading to more predictable trajectories and better model performance. On the other hand, the improved results may stem from the advantages of our roadside infrastructure-based setup, which benefits from a favorable viewpoint with reduced occlusion compared to the perspective of a self-driving car. This enhanced visibility likely contributes to the model’s ability to achieve lower ADE and FDE values, highlighting the potential of infrastructure-based systems in trajectory prediction tasks.

\begin{table}[!ht]
  \caption{HPNet performance on different datasets}\label{tab:hpnet-performance}
  \begin{center}
    \begin{tabular}{lcc}
      Dataset                           & minJointFDE (m)   & minJointADE (m)          \\\hline
      Argoverse                         & 1.0986            & 0.7612                   \\
      Interaction                       & 0.8231            & 0.2548                   \\
      \textbf{Work-zone Simulation}     & 0.3228            & 0.1327                   \\\hline
    \end{tabular}
  \end{center}
\end{table}

\subsection{Testing of Trajectory Prediction Model}

\begin{figure}
    \centering
    \includegraphics[width=0.95\linewidth]{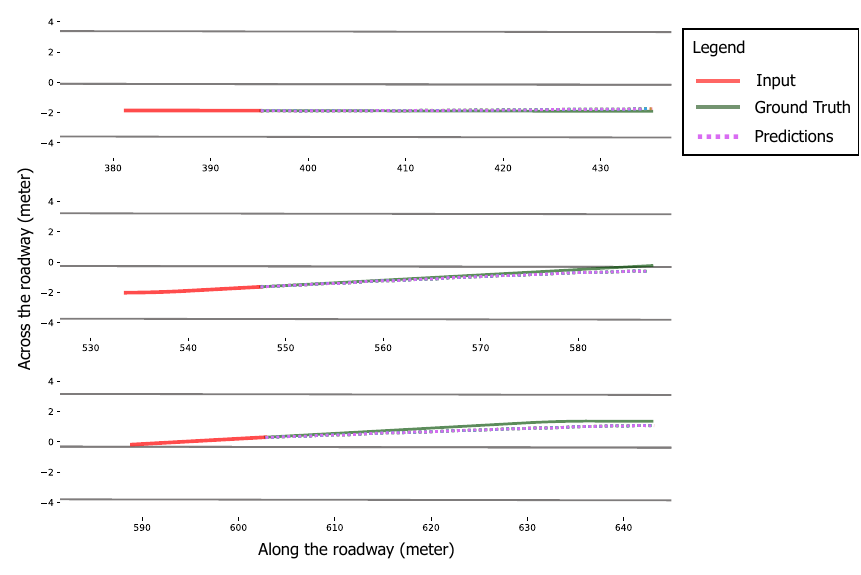}
    \caption{Predicted versus ground-truth trajectories for individual vehicles}
    \label{fig:track-prediction}
\end{figure}

Figure~\ref{fig:track-prediction} illustrates the predicted trajectories for selected individual vehicles over a 4-second horizon, plotted alongside ground-truth tracks. The figure covers both in-lane driving and lane changing cases. The predictions exhibit strong alignment with actual vehicle paths, capturing both the temporal evolution and spatial curvature of motion. The multi-modal nature of the prediction is evident in the closely clustered trajectories, which reflect subtle variations in potential driving behavior under the constraints of work zone geometry.


\subsection{Evalaution of Proactive Safety Warning Generation}
\label{sec:evaluation-warning-gen}

The conflict risk estimation framework discussed earlier in the subsection "Proactive Safety Warning Generation" was operationalized in the context of work zone safety analysis by applying it to the predicted vehicle trajectories generated over successive time horizons. A critical component of this application involved defining appropriate decision thresholds to classify interactions as potential conflicts and to issue proactive driver warnings. The first threshold, a spatial distance denoted $\delta_d = 7$, was selected as the cutoff below which a conflict event is considered likely. In work zone settings, lane widths are often reduced, merging tapers introduce lateral compression, and buffer space is minimal. A 7-meter minimum clearance between vehicle bounding boxes captures scenarios in which either longitudinal or lateral separation has diminished to a level that is no longer consistent with safe maneuverability. From a geometric standpoint, this threshold provides sufficient resolution to detect potential overlaps or near-miss trajectories, particularly once vehicles are aligned at oblique angles or operating in adjacent lanes under speed variability. Furthermore, the use of bounding-box geometry ensures that this 7-meter metric provides the shortest edge-to-edge distance, which is more safety-critical than centroidal spacing and aligns with real-world concerns about physical clearance and collision margins. To complement the spatial threshold, a secondary decision criterion was implemented based on probabilistic conflict scores. Specifically, when the predicted conflict probability exceeded a confidence level, a warning was generated and issued to the involved vehicles, indexed by their track IDs and associated prediction horizon. This probabilistic threshold was selected to balance sensitivity and specificity in warning generation. A value above 70\% indicates a high likelihood of physical convergence based on multi-step trajectory projections while still allowing for natural variability in motion prediction.  
\begin{figure}[H]
    \centering
    \includegraphics[width=0.9\linewidth]{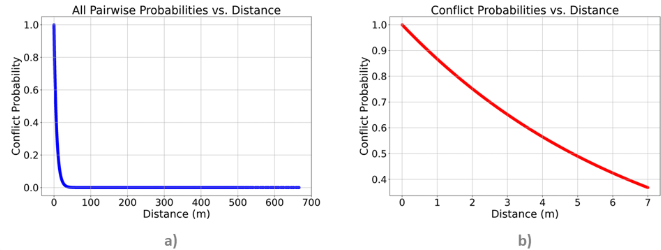}
    \caption{Conflict probability trends between two selected target vehicles across future prediction steps approaching a work zone area. (a) Full distribution of pairwise conflict probabilities at all predicted time horizons, illustrating how risk evolves over distance. (b) Filtered view showing only instances where the minimum bounding-box distance falls below the 7-meter critical threshold, highlighting high-risk interactions within safety-critical proximity.}
    \label{fig:warning-generation}
\end{figure}

The rationale for this second-level filter is twofold. First, it prevents the system from generating an excessive number of low-confidence alerts, which can degrade driver trust and lead to warning fatigue. Second, it allows the model to prioritize interactions with imminent safety consequences, focusing intervention efforts on cases with the highest inferred collision risk. This selective triggering makes sure that the warnings are not only accurate but also meaningful in operational contexts, where drivers or automated systems must respond decisively and with minimal distraction. Afterward, interactions falling below the 7-meter threshold are logged as conflicts, and those additionally exceeding the 70\% risk level are flagged for warning issuance. The output includes a time-indexed record of all flagged interactions, along with vehicle identifiers, conflict distances, and risk probability. To support this implementation, visualizations were developed to illustrate in Figure~\ref{fig:warning-generation} the empirical relationship between remaining distance and conflict probability across all evaluated vehicle pairs.

\begin{figure}[H]
    \centering
    \includegraphics[width=0.95\linewidth]{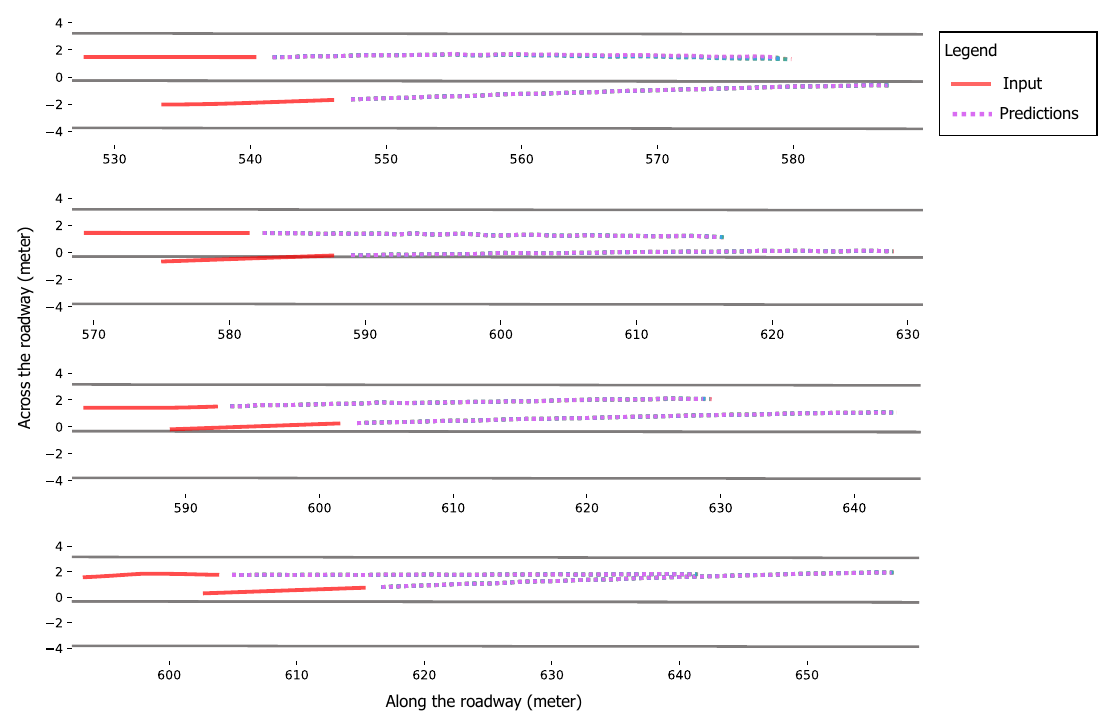}
    \caption{Predicted trajectory indicating potential conflict}
    \label{fig:stacked_conflict}
\end{figure}

Figure~\ref{fig:stacked_conflict} presents a potential conflict scenario where one vehicle performs an aggressive lane change maneuver. The visualization of the predicted trajectory shows a potential conflict before it takes place. The predicted trajectories are processed by the warning generation module and rightly flagged as a potential conflict. Early detection of conflict by the warning generation module utilizing the predicted trajectory is the basis for proactive safety, enabling the system to anticipate safety-critical events in advance. The use of HPNet's temporally consistent and spatially aware predictions enhances the reliability of these warnings in constrained and high-risk work zone environments.

\section{Conclusion}
This study presents a proactive safety framework for highway work zones, leveraging a predictive Digital Twin architecture enhanced by real-time sensing, HD maps, and temporally consistent trajectory prediction using the HPNet model. By adapting HPNet for the first time to an infrastructure-based deployment scenario, we demonstrate its effectiveness in forecasting vehicle trajectories with high accuracy in a work zone area. The co-simulation framework integrating SUMO and CARLA enables realistic data generation and testing, while the use of Lanelet2 maps allows for detailed topological encoding essential for trajectory forecasting in complex work zone environments. Evaluation results show that the adapted HPNet model achieves prediction accuracy on simulated work zone datasets, on par with benchmarks from established datasets, such as Argoverse and INTERACTION. Through a warning generation module based on vehicle bounding boxes and probabilistic conflict modeling, the system is capable of issuing alerts for potential vehicle conflicts. These findings underscore the potential of developing a predictive digital twin of a work zone area utilizing infrastructure-based sensing of the physical twin. This predictive digital twin system supports proactive safety in dynamic and constrained traffic environments. 

While the results indicate the efficacy of the proactive safety warning system, this study is limited by the scope of the test scenarios and the controlled simulated environment in which the evaluation was conducted. Future work will extend this framework to real-world deployments and incorporate cooperative behavior modeling through vehicle-to-everything communication for personalized safety warning generation to further enhance safety outcomes.

\section{Funding}
This material is based upon work supported by the Federal Motor Carrier Safety Administration under grant FM-MHP-0792.

\section{Acknowledgments}
Any opinions, findings, and conclusions or recommendations expressed this publication are those of the author(s) and do not necessarily reflect the view of the Federal Motor Carrier Safety Administration and/or the U.S. Department of Transportation.


We used ChatGPT4o to help rephrase parts of our own writing to improve clarity and editorial quality.

\section{Athors Contribution}
\textbf{Minhaj Uddin Ahmad:} conceptualization, coding, data collection, data analysis, methodology, and writing – original draft; \textbf{Alican Sevim:} data analysis and writing – original draft; \textbf{Mizanur Rahman:} conceptualization, data analysis, methodology, writing – original draft, and funding acquisition;  \textbf{David Bodoh:} coding and writing – review and editing;  \textbf{Sakib Khan, Li Zhao, Nathan Huynh and Eren Erman Ozguven:} conceptualization, writing – review and editing, and funding acquisition.

\newpage

\bibliographystyle{trb}
\bibliography{main}
\end{document}